\documentclass{snmp2003}

\begin{document}

\FirstPageHeading{Gavrilik}

\ShortArticleName{Quantum Groups and Cabibbo Mixing}

\ArticleName{Quantum Groups and Cabibbo Mixing}

\Author{Alexandre M. GAVRILIK~$^\dagger$}
\AuthorNameForHeading{A.M. Gavrilik}
\AuthorNameForContents{GAVRILIK A.M.}
\ArticleNameForContents{Quantum Groups and Cabibbo Mixing}
\Address{$^\dagger$~Bogolyubov Institute for Theoretical Physics,
       03143 Kiev, Ukraine}
\EmailD{omgavr@bitp.kiev.ua}

\Abstract{
Treating the issue of hadron masses and mass relations by the use
of quantum groups $U_q(su_n)$ taken as hadron flavor symmetries
suggests, at least in the case of baryons, a direct connection
of the deformation parameter $q$ with the Cabibbo angle.
We discuss possible manifestations of the Cabibbo mixing
implied by such connection, including unusual ones. }

\vspace{-2mm}

\section{Introduction}

The standard model of fundamental particles and forces
incorporates the important concept of quark mixing      \cite{Gavrilik:CKM-W}
usually described by the CKM matrix which
in the Wolfenstein's form looks as
\[
V_{\rm CKM}=
\begin{pmatrix}
V_{ud}  & V_{us}  &  V_{ub}   \\
V_{cd}  & V_{cs}  &  V_{cb}   \\
V_{td}  & V_{ts}  &  V_{tb}
\end{pmatrix}
=
\begin{pmatrix}
1\!-\!\frac12\lambda^2  & \lambda  &   A\lambda^3 (\rho\!-{\rm i}\eta) \\
-\lambda        &  1\!-\!\frac12\lambda^2  &    A\lambda^2  \\
A\lambda^3 (1\!-\!\rho\!-{\rm i}\eta)   &  - A\lambda^2 &    1
\end{pmatrix} .
\]
When keeping the first two quark families, only the parameter
$\lambda\approx 0.22$ (the sine of Cabibbo angle
$\theta_{\rm C}$) persists.
Cabibbo mixing                                          \cite{Gavrilik:CKM-W}
is known to play basic role in describing weak 
decays of mesons and baryons by means of flavor 
changing (e.g., strangeness changing) quark currents.
Our goal is to discuss some other, unusual implications
of the Cabibbo mixing which are deduced on the base of
adopting instead of the flavor groups $SU(n)$ their quantum
counterparts $U_q(su(n))$.
Note that the idea of using                              \cite{Gavrilik:JPA}
in place of $SU(n)$
the quantum groups (or $q$-deformed algebras)
$U_q(su_n)$ and their representations                 \cite{Gavrilik:dri-ji}
in order to treat hadronic  flavor symmetries
is natural, easily justifiable and very fruitful,
leading to many interesting implications      \cite{Gavrilik:JPA,Gavrilik:NP}.
The most important one is the possibility
to link the $q$-parameter with the angle
$\theta_{\rm C}$, and this enables to find novel
connections for the concept of Cabibbo mixing.
Say, the earliest modified (improved) version of 
the Gell-Mann - Okubo (GMO) mass relation for 
the $SU(3)$ octet baryons $\frac12^+$ derived in a 
particular version                                       \cite{Gavrilik:LTK}
of 'diquark-quark' model (the LTK model), 
involves certain parameters characterizing diquark 
and separate quark.
The alternative improved version of GMO for octet baryons
is the extremely precise mass sum rule (MSR) obtained with
$U_q(su_n)$. It exploits fixed value of the
$q$-parameter linked in a simple way to $\theta_{\rm C}$,
and this fact allows to find direct connection, through
the fixed $q$, of the parameters of LTK diquark-quark
model to the angle $\theta_{\rm C}$.
Besides this, we discuss the connections of the $\theta_{\rm C}$
with such concepts as anyonic statistics parameter from the anyonic
picture, as intercepts of the two- and three-pion (-kaon) correlation
functions drawn in recent experiments on relativistic heavy ion
collisions.

\section{Baryon mass sum rules from $q$-deformation}

\subsection{Mass relations for octet and decuplet baryons}

Consider first $\frac12^+$ baryons from $SU(3)$ octet.
Using $U_q(su_n), \ n=2,3,4,$ and $U_q(u_{4,1})$ (the latter
playing the role of dynamical $q$-algebra), the $q$-deformed
mass relation for octet baryons
\begin{equation}                         \label{qGMO}
   [2]M_N+\frac{[2] M_{\Xi}}{[2]-1}
   = [3] M_{\Lambda }
+ \Bigl ( \frac{[2]^2}{[2]-1}-[3]
    \Bigr ) M_{\Sigma }
   +\frac{A_q}{B_q}\left( M_{\Xi } + [2] M_N -
   [2]M_{\Sigma } - M_{\Lambda } \right) \
 \end{equation}
was obtained                                       \cite{Gavrilik:breg-uzh}.
The pair of polynomials  $A_q, B_q$ (of the $q$-deuce
$[2]\equiv[2]_q\!=\!q\!+\!q^{-1}$), with non-overlapping 
sets of zeros, results from calculation in a particular 
dynamical representation (irrep) of $U_q(u_{4,1})$; different
irreps lead to differing pairs $A_q, B_q$ in (\ref{qGMO}).
Each $A_q$ possesses the mandatory factor $([2]_q\!-\!2)$,
i.e. 'classical' zero $q=1$, and nontrivial zeros.
As particular cases, the $q$-analog (\ref{qGMO})
yields the familiar                                    \cite{Gavrilik:8fold}
Gell-Mann--Okubo (GMO) mass relation
$ M_N+M_\Xi=\frac32 M_\Lambda+\frac12M_\Sigma$
in the 'classical' limit $q=1$,  and two new
mass sum rules of improved accuracy           \cite{Gavrilik:NP,Gavrilik:GI}
\begin{eqnarray}
 & \hspace{-1.1mm} M_N+\frac{1\!+\!\sqrt{3}}{2}M_\Xi=
                    \frac{2 M_\Lambda}{\sqrt{3}}+
\frac{9\!-\!\sqrt{3}}{6}M_\Sigma     \hspace{12mm} (0.22\%) \ \ \
                                                        \label{q6} \\
 & \hspace{-6.7mm} M_N+\frac{M_{\Xi}}{[2]_{q_7}\!-\!1}=
\frac{M_{\Lambda}}{[2]_{q_7}\!-\!1}+M_{\Sigma}
         \hspace{21mm}  (0.07\%)  \hspace{-0.5mm}
                                                       \label{q7}
\end{eqnarray}
which correspond to $A_q$ and $\tilde{A}_q$
with respective zeros $q_6=e^{{\rm i}\pi/6}$ and $q_7=e^{{\rm i}\pi/7}$.
Among the admissible irreps of $U_q(u_{4,1})$ or its 'compact'
counterpart $U_q(u_5)$ there exist infinite series of irreps
producing infinite set of MSRs numbered by integer
$m$ $\ (6\le m<\infty )$, each given by               (\ref{qGMO})
with $q_m$ put for $q$, where $q_m=e^{{\rm i}\pi/m}$
guarantees vanishing of $\frac{A_q}{B_q}$.
Each MSR in this set {\it agrees better with data} than the
standard GMO one.
Thus, a 'discrete choice' (instead of usual fitting)
becomes possible; the $q$-polynomial $A_q$
due to zero $q_m$ serves as {\it defining} polynomial
for the corresponding MSR.
It is the value $q_7=e^{{\rm i}\pi/7}$ and the MSR (\ref{q7})
that provides the best choice.

For decuplet baryons ${\frac32}^+$, 1st order $SU(3)$
breaking yields well-known                             \cite{Gavrilik:8fold}
equal spacing rule (ESR) for isoplets
from ${\bf 10}$-plet.
However,
$M_{\Sigma^*}{-}M_{\Delta},$ $ M_{\Xi^*}{-}M_{\Sigma^*}$
and $M_{\Omega}{-}M_{\Xi^*}$ significantly deviate from ESR:
$152.6~MeV\leftrightarrow 148.8~MeV\leftrightarrow 139.0~MeV$.
The other known relation                     \cite{Gavrilik:oku,Gavrilik:LTK}
\begin{equation}
(M_{\Sigma^{*}}-M_{\Delta}+M_{\Omega}-M_{\Xi^{*}})/2
= M_{\Xi^{*}}-M_{\Sigma^{*}}                         \label{dec}
\end{equation}
accounts 1st, 2nd orders in $SU(3)$-breaking and holds only
slightly better than the ESR.

On the contrary, the use of $q$-algebras $U_q(su_n)$
gives nice improvement.
Evaluation of decuplet masses with particular irreps
of the dynamical  
$U_q(u_{4,1})$ yields the
$q$-deformed mass relation                                \cite{Gavrilik:GKT}
\begin{equation}
({M_{\Sigma^*}-M_{\Delta}+M_{\Omega}-M_{\Xi^*} })/
    ( { 2\cos\theta_{\bf 10}})
= M_{\Xi^*}-M_{\Sigma^*} ,     \hspace{18mm}
q=\exp({\rm i}\theta_{\bf 10})  .              \label{q-dec}
\end{equation}
This mass relation is
{\em universal} - it results from any admissible irrep
(containing $U_q(su_3)$-decuplet embedded in
{\em 20}-plet of $U_q(su_4)$ ) of the dynamical $U_q(u_{4,1})$.
With empirical masses, eq.            (\ref{q-dec})
holds remarkably for $\theta_{\bf 10}\simeq\frac{\pi }{14}$.
It is argued that
$\theta_{\bf 10}=\theta_{\rm{C}}$, see                    \cite{Gavrilik:NP}
and also next subsection.

\subsection{Linking the $q$-parameter to the Cabibbo angle}

We compare the generalization                          \cite{Gavrilik:oku70}
$f_\pi^2 m_\pi^2 + 3 f_\eta^2 m_\eta^2 = 4 f_K^2 m_K^2$
of GMO-formula $ m_\pi^2 + 3 m_\eta^2 = 4 m_K^2$
for pseudoscalar (PS) mesons, with
${f^{2}_\pi} + {3} {f^{2}_\eta} = {4}{f^{2}_K}$  imposed,
and our $q$-deformed analog
\begin{equation}
m_\pi^2 +  
           [3]_q^{-1}{(2~[2]_q - [3]_q)} m_{\eta_8}^2
= 2~[2]_q {(2~[2]_q - [3]_q)^{-1}} m_K^2 ,
         \hspace{12mm}     [3]_q=[2]_q^2-1 ,
\end{equation}
(found in                                                 \cite{Gavrilik:JPA})
of PS-mesonic GMO,
with duly fixed $q=q_{\rm{PS}}$ and
{\it physical} $\eta$-meson put in place of $\eta_8$
(that is, no need for explicit mixing).
We find, using
${\xi}_{\pi,K}\equiv({4 f^2_K/f^2_\pi})^{-1}$, that
\[
{f_K^2}/{f_\pi^2} \
      \leftrightarrow \  {\frac12 [2]}/({2[2]\!-\![3]}),
\ \ \ \ \ \ \ \ \
[2]_{\pm}=\!1\!-\!{\xi}_{\pi,K}\pm
({\bigl(1\!-\!{\xi}_{\pi,K}\bigr)^2\!+\!1})^{1/2} .
\]
Since the ratio $f_K/f_\pi$ is expressible 
through the Cabibbo angle (see e.g.,                  \cite{Gavrilik:oakes}),
we infer: the deformation parameter 
$q_{\scriptstyle{PS}}$ is directly connected 
with the Cabibbo angle.

In another way of reasoning, we use the result of       \cite{Gavrilik:isaev}
where the Lagrangian for quantum-group valued
gauge field analog of the Weinberg - Salam (WS)
model was constructed and the relation
$ \tan\theta = h(q)\equiv(1-q^2)/(1+q^2)$ found.
It provides proper mixing of the $U(1)$-component
$B_\mu$ and the nonabelian component $A^3_\mu$.
Physical photon $\tilde{A}_\mu$ and $Z$-boson of WS model
appear through the Weinberg angle
$\theta_{\rm {W}}=\theta=\arctan h(q)$.
At $\theta = 0$ the potentials $B_\mu$ and $A^3_\mu$
get unmixed, but the $q$-deformation with $\theta\ne 0$
provides mixing inherent in the WS model.
So, the mixing of (electro)weak gauge fields is
adequately modelled by the $q$-deformation.
Due to the latter, the weak angle and the $q$-parameter
are  explicitly linked.
The relation                                            \cite{Gavrilik:palle}
$\theta_{\rm W}
= 2 (\theta_{12} + \theta_{23} + \theta_{13})
$,
on the other hand, connects $\theta_{\rm {W}}$ with
the Cabibbo angle $\theta_{12}\equiv \theta_{\rm{C}}$
(we neglect the 3rd family's $\theta_{13}, \theta_{23}$).
Thus, the apparently different mixing angles, in the
{\it bosonic} (interaction) and in the {\it fermionic} (matter)
sectors of the electroweak model, are related.

We conclude that {\em the Cabibbo angle is linked with}
$q$-{\em parameter} of a quantum-group (or $q$-algebra)
based symmetry structure applied in the fermion sector.
The explicit connection is remarkably simple:
$
\theta_{\bf 10}=\theta_{\rm{C}}, \
\theta_{\bf 8}=2~\theta_{\rm{C}}
$.
With   $\theta_{\bf 8}=\frac{\pi}{7}$ this suggests
the exact value $\frac{\pi}{14}$
for $\theta_{\rm{C}}$ .

\subsection{Nonpolynomiality in $SU(3)$-breaking and Michel's statement}

The universality of $q$-analog (\ref{q-dec}) concerns all
admissible irreps of the {\it 'compact'} dynamical $U_q(su_5)$, too.
Say, calculation in the dynamical  irrep $\{ 4 0 0 0 \}$ of
$U_q(su_5)$  yields
$M_{\Delta}=M_{\bf 10} +  \beta ,\ $
$M_{\Sigma^*}=M_{\bf 10} + [2] \beta +  \alpha ,\ $
$M_{\Xi^*}=M_{\bf 10} + [3] \beta + [2] \alpha ,\ $
$M_{\Omega}=M_{\bf 10} + [4] \beta + [3] \alpha ,\ $
from which                                 (\ref{q-dec}) stems.
With hypercharge $Y$, all the decuplet
masses $M_{D_i} \equiv M\bigl( Y(D_i) \bigr )$
are comprised by single formula
\begin{equation}
M_{D_i}{=}M_{\bf 10} {+} {\alpha} [1\!-\!Y(D_i)]_q
             + {\beta} [2\!-\!Y(D_i)]_q                \label{Y-dec}
\end{equation}
of explicit $Y$-dependence.
The limit $q\to 1$ reduces it to
$M_{D_i} = \tilde{M}_{\bf 10} + a~Y\!(D_i)$
with $ a =\!-\!\alpha\!-\!\beta ,\ $
$\tilde{M}_{\bf 10} = M_{\bf 10}\!+\!\alpha\!+\!2\beta$,
i.e., to linear dependence on hypercharge
(or strangeness $S=Y\!-\!1$).

Formula (\ref{Y-dec}) involves {\it highly nonlinear dependence}
of mass on hypercharge: for decuplet, $Y$ alone causes $SU(3)$-breaking.
Since for any $N$ its $q$-number 
is
$[N]_q=(q^N-q^{-N})/(q-q^{-1})=
q^{N-1}+q^{N-3}+\ldots +q^{-N+3}+q^{-N+1}$ ($N$ terms),
this shows exponential $Y$-dependence of masses.
Such high nonlinearity makes           (\ref{q-dec}) and (\ref{Y-dec})
crucially different from the result       (\ref{dec})
of traditional treatment accounting linear
and quadratic effects in $Y$.

For octet baryon masses, {\em nonpolynomiality}
in $SU(3)$-breaking effectively accounted                 \cite{Gavrilik:GI}
by the model is embodied in the expressions
for isoplet masses, with explicit dependence on  
$Y$ as well as isospin $I$, through $I(I+1)$.
Matrix elements contributing to octet baryon masses
contain e.g., the terms
$\left({[Y/2]_q[Y/2\!+\!1]_q-[I]_q[I\!+\!1]_q}\right)$  or
$\left({[Y/2-1]_q[Y/2-2]_q-[I]_q[I+1]_q}\right)$,
with multipliers depending on irrep labels
$m_{\scriptstyle{15}},m_{\scriptstyle{55}}$, that show
explicit dependence on hypercharge and on the $q$-analog
$[I]_q[I+1]_q$ of $SU(2)$ Casimir.
The $q$-bracket $[n]_q$ means
$[n]_q=\frac{\sin(nh)}{\sin(h)}$, $ q\!=\!\exp({\rm i}h)$,
so we see that octet baryon masses depend on hypercharge $Y$
and isospin $I$ (hence, on $SU(3)$-breaking effects) also
in highly nonlinear - {\it nonpolynomial} - fashion.

We note finally that the conclusion made in the
preceding subsection about the ability to find, due to
the link $q\leftrightarrow \theta_{\scriptstyle{C}}$,
the exact value of Cabibbo angle is in accord
with Michel's statement                                  \cite{Gavrilik:mich}
that only account of higher-order breaking effects
enables gaining of such result.

   \section{ Cabibbo angle  and the anyonic statistics parameter}

From $N$ sorts of lattice fermions
$c_i({\bf x})$, $\ c_i^{\dagger}({\bf x})$, $\ i=1,...,N,$
with usual lattice anticommutation relations (ACRs),
using the lattice angle functions                       \cite{Gavrilik:lerda}
 $\theta_{\gamma}({\bf x},{\bf y})$ and
$\theta_{\delta}({\bf x},{\bf y})$ for two opposite
($\gamma$- and $\delta$-) types of cuts, the related
ordering on the lattice (${\bf x} > {\bf y}$ or
$ {\bf y} > {\bf x}$),  and the two types of
statistical operators $K_i({\bf x}_{\gamma})$ and
$K_i({\bf x}_{\delta})$,
\begin{equation}                              \label{K-oper}
K_j({\bf x}_{\gamma})=  \exp\bigl( {\rm i} \nu
    \sum_{{\bf y}\ne{\bf x}}\theta_{\gamma}({\bf x},{\bf y})
c_j^{\dagger}({\bf y})c_j({\bf y}) \bigr) , \ \ \ \
K_j({\bf x}_{\delta})=  \exp\bigl( {\rm i} \nu
    \sum_{{\bf y}\ne{\bf x}}\theta_{\delta}({\bf x},{\bf y})
c_j^{\dagger}({\bf y})c_j({\bf y}) \bigr) ,
\end{equation}
one defines                                             \cite{Gavrilik:lerda}
the anyonic oscillators
$
a_i({\bf x}_{\gamma})= K_i({\bf x}_{\gamma})c_i({\bf x}) \
$
and
$\ a_i({\bf x}_{\delta})= K_i({\bf x}_{\delta})c_i({\bf x})
$
involving {\em the anyonic statistics parameter} $\nu$.
From this definition and ACR's for lattice fermions,
the relations of permutation for anyonic oscillators
then follow. Some of them, e.g.
\[
a_i({\bf x}_{\gamma})a_i({\bf y}_{\gamma})
+q^{-{\rm sgn}({\bf x}-{\bf y})}
a_i({\bf y}_{\gamma})a_i({\bf x}_{\gamma}) =0,  {}     \ \ \ \ \ \
      a_i({\bf x}_{\gamma})a_i^{\dagger}({\bf y}_{\gamma})
       +q^{{\rm sgn}({\bf x}-{\bf y})}
          a_i^{\dagger}({\bf y}_{\gamma})a_i({\bf x}_{\gamma}) = 0,
\]
involve the deformation parameter $q$ connected with
the statistics parameter $\nu$ of (\ref{K-oper}) as
$\ q=\exp({\rm i}\pi\nu)$).
The generating elements $A_{j,j+1}$, $A_{j+1,j}$ and $H_j$
of the quantum algebra $U_q(su_N)$ are realized bilinearly
through anyonic oscillators $a_i({\bf x}_{\gamma})$,
$a_i^{\dagger}({\bf y}_{\gamma})$ and shown to satisfy   \cite{Gavrilik:lerda}
the defining relations                                 \cite{Gavrilik:dri-ji}
of the quantum algebra $U_q(su_N)$.
 Dual realization in terms of  $a_i({\bf x}_{\delta})$,
$a_i^{\dagger}({\bf y}_{\delta})$ is also valid.
 Then, in anyonic realization of $U_q(su_N)$,
both hadron mass operator $\hat{M}$ and basis vectors
for hadronic irreps are explicitly constructed            \cite{Gavrilik:GI2}.
   Say, for the irrep $\{4000\}$ of 'dynamical' $U_q(su_5)$,
in accordance with the chain of $q$-algebras
$U_q(su_3)\subset U_q(su_4)\subset U_q(su_5)$ and respective
chain of irreps $[30]\subset [300]\subset \{4000\}$,
all basis state vectors
$
|n_1n_2n_3n_4\rangle \equiv
a_{n_1}^{\dagger}({\bf x}_{1\gamma})
a_{n_2}^{\dagger}({\bf x}_{2\gamma})
a_{n_3}^{\dagger}({\bf x}_{3\gamma})
a_{n_4}^{\dagger}({\bf x}_{4\gamma})|0\rangle
$
of baryons $\frac32^+$ are constructed by acting
with lowering generators upon the highest weight vector.
 E.g., for isoquartet baryon $|\Delta^{++}\rangle$ we get
$
|\Delta^{++}\rangle =
[4]^{-1/2}(|5111\rangle + q^{-1}|1511\rangle +
q^{-2}|1151\rangle + q^{-3}|1115\rangle),
$
and similarly for $|\Sigma^*\rangle$, $|\Xi^*\rangle$,
$|\Omega^-\rangle$.
The dual basis vectors
${|\Delta^{++}\!\tilde\rangle}$ etc., are also given.
Masses $M_{D_i}$ of baryons $D_i$ in the dynamical
$U_q(su_5)$-irrep $\{4000\}$ are calculated with
mass operator $\hat{M}$ as
$M_{D_i}={\tilde\langle D_i|}\hat{M}|D_i\rangle $
to yield:
$M_{\Delta}=M_{\bf 10}+\beta$,
$M_{\Sigma^*}=M_{\bf 10}+[2]_q\alpha+[2]_q\beta$,
$M_{\Xi^*}=M_{\bf 10}+[2]_q^2\alpha+[3]_q\beta$,
and
$M_{\Omega^*}=M_{\bf 10}+[2]_q[3]_q\alpha+[4]_q\beta$,
from which the relation (\ref{q-dec}) follows.
This proves applicability                                 \cite{Gavrilik:GI2}
of quantum algebras and their irreps for deriving
hadron mass relations by {\it using their anyonic realization}.

Since $\theta_{\bf 10}=\theta_{\scriptstyle{C}}$,
{\em we have the connection: Cabibbo angle}
$\leftrightarrow$ {\em anyonic statistics parameter} $\nu$.

\section{ Diquark-quark model parameters and the Cabibbo angle }

The LTK diquark-quark model                               \cite{Gavrilik:LTK}
uses, besides the $SU(3)$ invariant masses
$m_t$,  $m_s$ and $m_q$ of the $SU(3)$ diquark triplet,
diquark sextet, and 3rd quark triplet (so
the subscripts $'t'$, $'s'$, $'q'$), also the
mass parameters $\delta_t$,  $\delta_s$ and $\delta_q$
measuring $SU(3)$ violation in the respective multiplets.

The improved form of GMO obtained in the LTK model
is
\begin{equation}                           \label{C-LTK}
{\textstyle \frac32} m_\Lambda
+ {\textstyle \frac12} m_\Sigma - m_N - m_\Xi
= {\cal C}_{\rm LTK}    \equiv \mu_s \bigl (
3 \xi_{ts}^2 + 18 \xi_{ts} - 13 \bigr )     \hspace{0.6mm}
\end{equation}
where
$\xi_{ts} = \frac{\delta_t-\delta_q}{\delta_s-\delta_q}$,
$\ \mu_s = V_0  \gamma_s^2$,
$\ \gamma_s= \frac
{(\delta_s -\delta_q)}{6V_0}$.
The  $\mu_s$ must be positive being also involved         \cite{Gavrilik:LTK}
in the decuplet mass combination:
$8 \mu_s = 2 m_{\Xi^*} - m_{\Omega} - m_{\Sigma^*} > 0.$
As result, the LTK model gave the value
${ \xi^{\rm LTK}_{ts}=-3 }$ which implied:
$ \delta_t$ must be respectively greater (less) 
than $\delta_q$ with the parameter $\delta_s$ being less (greater) 
than $\delta_q$.
However, the value $\xi^{\rm LTK}_{ts}=-3$ contradicts data 
as it renders the r.h.s. of                       (\ref{C-LTK})
to be negative thus correcting the GMO
{\em in wrong direction}.

On the other hand, our $q$-MR (\ref{q7})
for which $q_7=\exp({\rm i}\frac{\pi}{7})$, \
$\frac{\pi}{7}{=}2\theta_{\rm C} $,
can be rewritten as
\begin{equation}                                 \label{C-q7}
{\textstyle\frac32} m_\Lambda\!+\!{\textstyle\frac12}
m_\Sigma\!-\!m_N\!-\!m_\Xi
= {\cal C}_{q_7}
\equiv
\frac{2-[2]_7} {{[2]_7-1}}
(m_\Xi - m_\Lambda)\!-\!{\textstyle\frac12} (m_\Sigma - m_\Lambda).     \ \ \
\end{equation}
Comparison of the two improved versions
(\ref{C-LTK}), (\ref{C-q7}) leads to the relation          \cite{Gavrilik:GK}
\begin{equation}                            \label{LTK-q7}
 ([2]_7 - 1)^{-1} =
\frac{ 9 \tilde\xi_{ts}^2 - 6 \tilde\xi_{ts} + 5 }
     { 4 (\tilde\xi_{ts}^2 - 6 \tilde\xi_{ts} + 5)} ,
 \ \ \ \ \ \ \
\tilde\xi_{ts}=\frac{\tilde\delta_t-\tilde\delta_q}
                    {\tilde\delta_s-\tilde\delta_q} ,
\end{equation}
which connects the value $q_7 = \exp (i\frac{\pi}{7})$ of
$q$-parameter in the $q$-GMO (\ref{C-q7}) with the ratio
$\xi_{ts}$  of the LTK model.                               %
Remark that the values of $\tilde\delta_t$,
$\tilde\delta_s$ and $\tilde\delta_q$ in this relation
are {\em understood as the optimized ones} reflecting,
through the found connection, {\em all-order account} 
by (\ref{C-q7}) of $SU(3)$ symmetry breaking in 
octet baryon masses - just this fact is denoted by tildas
over $\delta_t$, $\delta_s$, $\delta_q$.

Since $[2]_7=2\cos{\frac\pi7}\approx 1.80194$,
solving the relation  (\ref{LTK-q7}) yields
the values $ \xi_{ts}^{(+)} \approx 0.741 $
and $\xi_{ts}^{(-)} \approx - 6.705 .$
By the very derivation, $\xi^{(\pm)}_{ts}$ should
guarantee  the validity of sum rule (\ref{C-LTK})
to within  $0.07 \% .$
The both values differ essentially from the
value $\xi^{\rm LTK}_{ts}=-3$ of LTK model,
{\em providing proper positive correction to} GMO,
see the above comment about $\xi^{\rm LTK}_{ts}$.
Moreover, being positive, our $\xi_{ts}^{(+)}$
is not only well acceptable phenomenologically,
but also reflects {\em qualitatively} different, 
than $\xi_{ts}^{(-)}$ and $\xi^{\rm LTK}_{ts}$,
physical situation.
 Namely, the mass parameter $\tilde\delta_t$
is greater (less) than $\tilde\delta_q$ when
the mass parameter $\tilde\delta_s$
is greater (less) than $\tilde\delta_q$, since
\[
\tilde\delta_t -\tilde\delta_q = 0.741 (\tilde\delta_s -\tilde\delta_q)
\ \ \ \ \ \ \ \ {\rm i.e.} , \ \ \ \ \ \
\tilde\delta_t -\tilde\delta_s = 0.259 (\tilde\delta_q -\tilde\delta_s) .
\]
Thus, the inequalities for the (optimized) parameters
of LTK model should be
\begin{equation}                                    \label{ineq}
{\rm either}  \ \ \ \ \ \ \ \
\tilde\delta_q  > \tilde\delta_t  > \tilde\delta_s
           \hspace{12mm}
\hbox{ or}
           \hspace{12mm}
\tilde\delta_s  > \tilde\delta_t > \tilde\delta_q .
\end{equation}
Since $\theta_{\bf 8}=\frac{\pi}{7}=2~\theta_{\rm C} $,
or   $\ \theta_{\rm C}=\frac{\pi}{14}$,
we finally arrive                                          \cite{Gavrilik:GK}
at the formula yielding
{\em direct link} of the Cabibbo angle to the
(optimized) parameters of the LTK diquark-quark model:
\begin{equation}                                 \label{LTK-t'c}
\cos{2 \theta_{\rm C}}
= \frac{13\tilde\xi^2_{ts}-30\tilde\xi_{ts}+25}
         {2(9{\tilde\xi}^2_{ts}-6{\tilde\xi}_{ts}+5)},
   \hspace{14mm}
\tilde\xi_{ts}\equiv\frac{\tilde\delta_t-\tilde\delta_q}
                         {\tilde\delta_s-\tilde\delta_q} . \ \
\end{equation}
This connection of the Cabibbo angle with the parameters
of diquark-quark model, clearly, could not be found
without eq. (\ref{q7}) (or eq. (\ref{C-q7}))
and the relation (\ref{LTK-q7}) deduced on its base.

\section{ Cabibbo mixing in multiparticle correlations?}

The model of ideal gas of $q$-bosons based on the algebra of
$q$-deformed oscillators of
Arik-Coon (AC) or
Biedenharn-Macfarlane (BM) type                         \cite{Gavrilik:AC-BM},
may be used to describe                                   \cite{Gavrilik:AGI}
unusual behaviour of 2-particle correlations
of identical pions or kaons produced in
relativistic heavy ion collisions.
The approach yields explicit expressions and clear
predictions                                               \cite{Gavrilik:AGP}
for the intercept $\lambda$
(dependent on the temperature, particle mass,
pair mean momentum, {\it and the deformation parameter} $q$).

Physical observables  are evaluated as thermal averages
$\langle A \rangle ={\rm Sp}(A\rho )/{\rm Sp}(\rho)$,
$\rho = e^{-\beta H}$, where the Hamiltonian is
$H={\sum}\omega_i N_i$ and $\beta=1/T$.
With $b^\dagger_i b_i=[N_i]_q$ and $[2]_q=2\cos\theta$,
the $q$-deformed distribution function for the BM-type
$q$-bosons results (see e.g.                          \cite{Gavrilik:AGI})
as
\vspace{-0.3mm}
\begin{equation}
\langle b_i^\dagger b_i \rangle = 
({e^{\beta\omega_i}-1})/
({e^{2\beta\omega_i}-2\cos\theta~e^{\beta\omega_i}+1}).    \label{21}
\end{equation}
\vspace{-0.3mm}
At $\theta{=}0$ (or $q{=}1$), it yields Bose-Einstein (BE) distribution,
as $q{=}1$ recovers usual bosonic commutation relations.
Deviation of $q$-distribution (\ref{21}) from the quantum
BE distribution is seen to tend towards the
Boltzmann one (reduced quantum statistical effects).
For kaons, whose mass $m_K>3m_{\pi}$, analogous curve
gets closer (than pion's one) to the BE distribution.
Note that for AC-type $q$-bosons, the $q$-distribution
looks more simple:
$
\langle b_i^\dagger b_i \rangle = (e^{\beta\omega_i}-q)^{-1} .
$

\subsection{ Intercept $\lambda$ of two-pion (two-kaon) correlations
             and Cabibbo angle}

To obtain explicitly the intercept $\lambda$ of
two-particle correlations one calculates the two-particle
distribution $\langle b^\dagger b^\dagger b b\rangle$ and
normalizes it by $\langle b^\dagger b \rangle^2$.
The result, see                                           \cite{Gavrilik:AGI},
for AC-type $q$-bosons reads
$\lambda=
q-\frac{q(1-q^2)}{e^{\omega/T}-q^2}$, $\ -1\le q\le 1$,
and for BM-type $q$-bosons, with
${\cal F}(\beta\omega)\equiv\cosh(\beta\omega)$,
it is
\begin{equation}
\lambda+1\equiv\langle b^\dagger b^\dagger b b\rangle /
(\langle b^\dagger b \rangle)^2 =      
\frac{ 2\cos\theta~({\cal F}(\beta\omega)\!-\!\cos\theta)^2 }
{ ({\cal F}(\beta\omega)\!-\!1)
({\cal F}(\beta\omega)\!-\!2\cos^2\theta+1) }.
                                                    \label{22}
\end{equation}

With $\omega=(m^2+{\bf K^2})^{1/2}$,
asymptotically at large mean momentum of pion (kaon) pair
and fixed temperature, the intercept $\lambda$ tends
to the constant $\lambda^{\rm AC}_{\rm asymp}=q\ $
for the AC-type $q$-bosons ($q$ real), and for the BM-type
$q$-bosons ($q$ is a phase factor) to the constant
\begin{equation}
\vspace{-0.6mm}
\lambda^{\rm BM}_{\rm asymp}=2\cos\theta-1, \ \ \ \ \ \ \ \
                       \theta= -{\rm i}~{\ln}~q .  \label{23}
\vspace{-0.6mm}
\end{equation}
As suggested in                            \cite{Gavrilik:AGP,Gavrilik:laue},
correlations of pions and kaons are characterized
by the same value of $q$ (a kind of universality).
Then, experimentally there should be a tendency of merging
$\lambda(\pi)$ and $\lambda(K)$ at large enough mean momenta:
$ \lambda_{\rm asymp}(\pi)=\lambda_{\rm asymp}(K)$.
Recent RHIC/STAR data give (see discussion in          \cite{Gavrilik:laue})
the values $\lambda_1(\pi^-)$, $\lambda_2(\pi^-)$
and $\lambda_3(\pi^-)$ for $\pi^-$-intercept, as averaged
over three intervals of transverse momenta ${\bf K}_t$
and also over rapidity $y$,  $\ -0.5 \le y \le 0.5$.

    {
   \begin{figure}[t]
\begin{center}
\vspace{0.7cm}
\hspace{-2mm}   {
\epsfig{file=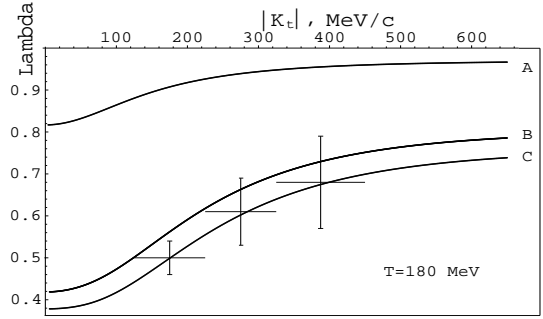,height=3.6cm,width=6.8cm,angle=0}      }
\caption  { The intercept $\lambda $ of two-pion correlation
function versus transverse momentum $|{\bf K}_t|$ at fixed $T=180$ MeV
and fixed  $q=\exp(i\theta)$ :
 \  A) $\theta = 10^\circ$,
 \  B) $\theta =  25.7^\circ\ $ (i.e., $2 \theta_{\rm C}$),
 \  C) $\theta = 28.5^\circ$ .}
\vspace{-0.3cm}
\end{center}
\end{figure}
  }

In Fig.1, we show three curves for the intercept $\lambda$
which correspond to fixing in (\ref{22}) certain values
of the deformation angle $\theta$, along with the data
$\lambda_j(\pi^-)$, $j{=}1,2,3$ (with error bars).
The temperature is $T=180$ MeV for all curves.
We find remarkable agreement with data for the curve C
(i.e., $\theta=28.5^{\circ}$).
Another interesting case is the curve B at
$\theta=\frac{\pi}{7}=2\theta_{\rm C}\simeq 25.7^{\circ}$.
At $T=180$ MeV, the curve B agrees, within error bars,
with the points $\lambda_2(\pi)$ and $\lambda_3(\pi)$.
However, slightly higher effective temperature
$T\simeq 198$ MeV makes the curve marked by
$\theta\!=\!2\theta_{\scriptstyle{C}}$
respecting {\it all the three error bars}.
Among various mixing angles known for hadrons,
only the angle $2\theta_{\scriptstyle{C}}$ seems to be
relevant to the issue of intercept $\lambda(\pi)$.
Hence, it is tempting to suggest that just the angle
$2\theta_{\scriptstyle{C}}$ embodies the assumed
universality, to be seen in 2-particle correlations:
\begin{equation}
\lambda^{\rm BM}_{\rm asymp}(\pi)\vert_{\theta=\pi/7}=    \label{24}
\lambda^{\rm BM}_{\rm asymp}(K)\vert_{\theta=\pi/7}=
2\cos\frac{\pi}{7}{-}1\approx 0.80194.
\end{equation}
Insisting on the asymptotical coincidence
$\lambda_{\rm asymp}(\pi)\!=\!\lambda_{\rm asymp}(K)$,
we predict for kaons:
{\it the intercept of 2-kaon correlations
at any transverse momenta should obey:}
$\lambda(K) \lesssim 0.80194$.


\subsection{ Three-particle correlations of pions (viewed as $q$-bosons)
             and $\theta_{\rm{C}}$ }


Now consider higher order (or multi-particle)
correlations, first in the case of AC-type $q$-oscillators.
It is not difficult to derive                            \cite{Gavrilik:AdGa}
the following 3-particle monomode correlation 
function
\begin{equation}                                    \label{25}
\langle a_i^\dagger a_i^\dagger a_i^\dagger a_i a_i a_i\rangle =
\frac{(1+q)(1+q+q^2)}
{(e^{\eta_i}-q)(e^{\eta_i}-q^2)(e^{\eta_i}-q^3) }
\end{equation}
with $\eta_i\equiv \beta\omega_i $.
From this, and 1-particle distribution
$\langle  a_i^\dagger a_i \rangle = \frac{1}{e^{\eta_i}-q}$,
the intercept of 3-particle correlations for AC-type $q$-bosons
in monomode case, dropping the subscript "i", reads:
\begin{equation}                                 \label{26}
\lambda^{(3),{\rm AC}}+1=
\frac{\langle (a^\dagger)^3 a^3\rangle }
     {\langle a^\dagger a\rangle^3 }=
 \frac{(1+q)(1+q+q^2)(e^{\eta_i}-q)^2}
      {(e^{\eta_i}-q^2)(e^{\eta_i}-q^3)} ,  \ \ \ \ \ \ -1\le q\le 1 .
\end{equation}
The $n$-particle generalization 
of (\ref{25}) is obtained                               \cite{Gavrilik:AdGa}
in the form
\begin{equation}
\langle ( a_i^\dagger )^n  ( a_i )^n \rangle =    \label{27}
\frac{   \lfloor n\rfloor_q! }
{\prod_{r=1}^n (e^{\eta_i}-q^r) } \ ,  \ \ \ \ \ \
\lfloor m\rfloor_q\equiv\frac{1-q^m}{1-q}=1+q+q^2+...+q^{m-1} \
\end{equation}
where $\lfloor n\rfloor_q!=
\lfloor 1\rfloor_q\lfloor 2\rfloor_q \cdots\lfloor n\rfloor_q$.
Moreover, this result admits direct two-parameter 
extension                                              \cite{Gavrilik:AdGa}
to the model of $pq$-Bose gas, based on the 
use of so-called $pq$-oscillators, in the form
\begin{equation}
\langle ( A_i^\dagger )^n  ( A_i )^n \rangle =      \label{28}
\frac{   [\![n]\!]_{qp}! \ (e^{\eta_i}-1) }
{\prod_{r=o}^n ( e^{\eta_i} - q^r p^{n-r} ) } \ ,  \ \ \ \ \ \
[\![m]\!]_{qp}\equiv\frac{q^m-p^m}{q-p} \ .
\end{equation}
With the use of one-particle distribution
$\langle A^\dagger A\rangle =\frac{ e^w-1 }{(e^w-q)(e^w-p)}$
of $qp$-bosons (see also                                \cite{Gavrilik:daoud})
we obtain general formula                                \cite{Gavrilik:AdGa}
for the $qp$-deformed $n$-th order intercept
$\lambda^{(n)}_{q,p}$:
\begin{equation}
\lambda^{(n)}_{q,p} \equiv                         \label{29}
- 1 +
\frac{{\langle A^{\dagger n} A^{n} \rangle}}{{\langle
A^{\dagger} A \rangle}^{n}} =
- 1 + [n]_{qp}! \frac{(e^w-p)^n (e^w-q)^n}{(e^w-1)^{n-1}
\prod_{k=0}^n(e^w-q^{n-k}p^k)} .
\end{equation}
From this, the result (\ref{26}) for AC-type $q$-bosons
follows if $p=1$.
Similarly, putting $p=q^{-1}$ reduces to the important
case of BM-type $q$-bosons which for the 3-particle
intercept yields
\begin{equation}
\lambda^{(3),{\rm BM}} \equiv                           \label{30}
- 1 + \frac{{\langle a^{\dagger 3} a^{3} \rangle}}{{\langle
a^{\dagger} a \rangle}^{3}} =
- 1 + \frac{ [2]_{q}[3]_{q} (e^{2w}-2\cos\theta e^{w}+1)^2 }
           { (e^w-1)^2  (e^{2w}-2\cos(3\theta) e^{w}+1)  } \
\end{equation}
(compare it with eq. (\ref{26})).
Again, like the intercept 
$\lambda^{(2),{\rm BM}}_{\rm asymp}$ in (\ref{23}), 
at large $w$ (large momenta or low $T$) we get 
$\lambda^{(3)}_{\rm asymp}$ dependent solely on the 
deformation parameter $q=\exp({\rm i \theta})$, i.e.
\begin{equation}
\lambda^{(3),{\rm BM}}_{\rm asymp} =
- 1 + [2]_q [3]_q  =                       \label{31}
- 1 + 2\cos\theta(2\cos\theta-1)(2\cos\theta+1)\ .
\end{equation}
To confront the results (\ref{29}),(\ref{31}) with
experimental data,  consider the combination             \cite{Gavrilik:r0}
\begin{equation}
r_0 =                                      \label{32}
{(\lambda^{(3)}-3\lambda^{(2)} ) }{(\lambda^{(2)} )^{-3/2}} .
\end{equation}
Then, the $r_0$ that follows from (\ref{22}) and (\ref{30})
(or $r_{0,{\rm asymp}}$ stemming from (\ref{23}) and (\ref{31})) 
is at fixed $T$ a decreasing function of $\theta$ 
for $0 \le \theta < \pi/3$.
What about the Cabibbo angle?
If we take the value $\theta=2\theta_{\rm C}$ of
deformation angle as we did in the preceding subsection
we get $r_0|_{\theta=2\theta_{\rm C}} \simeq 0.8955$.
The universality conjecture dictates now that  
{\it all posible values of} the $r_0$    (composed of 
intercepts $\lambda^{(2)}$ and $\lambda^{(3)}$) for 
either pions or kaons should respect the value
$0.8955$.
The presently available data                             \cite{Gavrilik:r0}
extracted in Pb-Pb and Au-Au collisions
seems to be yet insufficient to prove or disprove this assertion.

\section{ Conclusion }

The use of quantum groups (quantum algebras) in the context
of baryon phenomenology implies important fact that the
deformation parameter $q$ is linked very simply to
the basic (fermion) Cabibbo mixing angle $\theta_{\rm C}$.
In turn, this leads to unexpected connections for
$\theta_{\rm C}$ and for the whole concept of Cabibbo mixing,
e.g., with the parameters of diquark-quark model of baryons,
with the statistics parameter of anyonic picture, with
the experimentally testable intercept parameters of
multi-pion (-kaon) correlation functions, etc..
Of course, it would be highly desirable to (re)obtain
the considered connections of $\theta_{\rm C}$
in a more strict way, and we hope this will
be realized in a not very distant future.

\subsection*{Acknowledgements}

The author thanks  L.~Boya and J.~Beckers for
interesting discussions during the conference.


\LastPageEnding


\begin{thebibliography}{99}
\footnotesize

\bibitem{Gavrilik:CKM-W}  Cabibbo N.,
Unitary symmetry and leptonic decays.
{\em Phys. Rev. Lett.}, 1963, V.{10}, 531--533. \
     Kobayashi M. and Maskawa T.,
CP-violation in the renormalizable theory of weak interaction,
{\em Progr. Theor. Phys.}, 1973, V.{49}, N.2, 652--657. \
     Wolfenstein L.,
Parametrization of the Kobayashi-Maskawa matrix,
{\em Phys. Rev. Lett.}, 1983, V.{51}, N.21, 1945--1947.

\bibitem{Gavrilik:JPA}    Gavrilik A.M.,
$q$-Serre relarions in $U_q(u_n)$, $q$-deformed
meson mass sum rules,  and Alexander polynomials,
{\em J. Phys. A}, 1994, V.27, N.3, L91--L94.

\bibitem{Gavrilik:dri-ji}   Drinfeld V.,
Hopf algebras and the quantum Yang-Baxter equation,
Sov.Math.Dokl. {\bf 32} (1985) 254--258. \
         Jimbo M., Lett.Math.Phys. {\bf 10} (1985) 63--69.  \
Faddeev L.D., Reshetikhin N. and Takhtajan L.,
Quantization of Lie groups and Lie algebras,
{\em Leningrad Math. J.}, 1990, V.{1}, 193--225.

\bibitem{Gavrilik:NP}     Gavrilik A.M.,
Quantum algebras in phenomenological description of particle properties,
{\em Nucl. Phys. B (Proc. Suppl.)}, 2001, V.102/103, 298--305,\
    {\tt hep-ph/0103325}.

\bibitem{Gavrilik:LTK}    Lichtenberg D.B., Tassie L.J. and Keleman P.J.,
Quark-diquark model of baryons and SU(6).
{\em Phys. Rev.}, 1968, V.{167}, 1535--1542.

\bibitem{Gavrilik:breg-uzh}   Gavrilik A.M.,
in {\sl Symmetries in Science VIII} (Proc. Int. Conf., B.Gruber ed.)
 Plenum, N.Y., 1995, p. 109.      \\
Gavrilik A.M., Kachurik I.I. and Tertychnyj A.V.,
{\sl Kiev preprint} ITP-94-34E, 1994, \
   {\tt hep-ph/9504233}.

\bibitem{Gavrilik:8fold}   Gell-Mann M. and Ne'eman Y.,
The Eightfold Way, New York, Benjamin, 1964. \
Okubo S.,   $\varphi$-meson and unitary symmetry model,
{\em Phys.~Lett.}, 1963, V.{5}, 165--169.

\bibitem{Gavrilik:GI} Gavrilik A.M. and Iorgov N.Z.,
Quantum groups as flavor symmetries: account of
nonpolynomial SU(3)-breaking effects in baryon masses,
{\em Ukr. J. Phys.}, 1998, V.43, N.12, 1526--1533,\ {\tt hep-ph/9807559}.

\bibitem{Gavrilik:oku}      Okubo~S.,
Some consequences of unitary symmetry model,
{\em Phys.~Lett.}, 1963, V.{4}, 14--16.

\bibitem{Gavrilik:GKT}  Gavrilik A.M., Kachurik I.I. and Tertychnyj A.V.,
Baryon decuplet masses from the viewpoint of $q$-equidistance,
{\em Ukr. J. Phys.}, 1995, V.40, N.7, 645--649.

\bibitem{Gavrilik:oku70}   Okubo S.,
Test of quark models and asymptotic symmetry, in
Proc. Int. Conf. {"Symmetries and quark models"}
(editor R.Chand), New York, Gordon and Breach,1970, 59--79.

\bibitem{Gavrilik:oakes}   Oakes R.J.,
 SU(2)$\!\times\!$SU(2) breaking and the Cabibbo angle
{\em Phys.~Lett.}, 1969, V.{29}, 683--685.

\bibitem{Gavrilik:isaev}     Isaev A.P. and Popowicz Z.,
$q$-Trace for quantum groups and $q$-deformed Yang - Mills theory.
{\em Phys. Lett. B}, 1992, V.{281}, 271--278.

\bibitem{Gavrilik:palle}      Palle D.,
On the broken gauge, conformal and discrete symmetries
in particle physics, {\em Nuovo Cim. A}, 1996, V.{109}, 1535--1554.

\bibitem{Gavrilik:mich}   Michel L.,
On the dynamical breaking of SU(3),
in Proc. Fifth Coral Gables Conference {"Symmetry principles at high energy"}
(22-26 January, 1968), Benjamin, N.Y., 1968, 19--48.

\bibitem{Gavrilik:lerda}    Lerda A. and Sciuto S.,
Anyons and quantum groups,
{\em Nucl. Phys. B}, 1993, V.{401}, 613--637.

\bibitem{Gavrilik:GI2}  Gavrilik A.M. and Iorgov N.Z.,
Masses of decuplet baryons treated within anyonic realization
of the $q$-algebras $U_q({\rm su}_N)$,
{\em Ukr.~J.~Phys.}, 2000, V.45, N.7, 789--794,\ {\tt hep-ph/9912222}.

\bibitem{Gavrilik:GK}  Gavrilik~A.M. and Kachurik~I.I.,
Linking the parametetrs of diquark-quark model to Cabibbo angle,
{\em Ukr.~J.~Phys.}, 2003, V.48, N.6, 513--517,\ {\tt hep-ph/0301020}.

\bibitem{Gavrilik:AC-BM}   Arik~M. and Coon~D.D.,
Hilbert spaces of analytic functions and generalized coherent states,
{\em J.~Math.~Phys.}, 1976, V.{17}, 524--527.
Biedenharn~L.C.,
The quantum group SU$_q(2)$ and a $q$-analogue of the boson operators,
{\em J.~Phys. A.}, 1989, V.{22}, L873--L878.
MacFarlane~A.,
On $q$-analogs of the quantum harmonic oscillator and
the quantum group SU(2)$_q$,
{\em J.~Phys. A}, 1989, V.{22}, 4581--4588.

\bibitem{Gavrilik:AGI} Anchishkin~D.V., Gavrilik~A.M. and Iorgov~N.Z.,
Two-particle correlations from the $q$-boson viewpoint,
{\em Eur.~Phys.~J. A}, 2000, V.7, 229-238,\ {\tt nucl-th/9906034}.

\bibitem{Gavrilik:AGP} Anchishkin~D.V., Gavrilik~A.M. and Iorgov~N.Z.,
$q$-Boson approach to multiparticle correlations,
{\em Mod.~Phys.~Lett. A}, 2000, V.15, N.26, 1637--1646,\
{\tt hep-ph/0010019}.

\bibitem{Gavrilik:laue}  
Anchishkin~D.V., Gavrilik~A.M. and Panitkin~S.,
Transverse momentum dependence of intercept parameter $\lambda$
of two-pion (-kaon) correlation functions in $q$-Bose gas model,\
{\tt hep-ph/0112262}.

\bibitem{Gavrilik:AdGa}   Adamska~L.V. and Gavrilik~A.M.,
Multi-particle correlations in $qp$-Bose gas model,\
{\tt hep-ph/0312390}.

\bibitem{Gavrilik:daoud} Daoud~M. and Kibler~M.,
Statistical mechanics of $qp$-bosons in $D$ dimensions,
{\it Phys.~Lett. A}, 1995, V.206, N~1, 13--17.

\bibitem{Gavrilik:r0}  STAR Collaboration (Adams~J. et al.),
Three-pion HBT correlations in relativistic heavy ion collisions
from the STAR experiment,\  {\tt nucl-ex/0306028}.

\end{thebibliography}
\end{document}